# Geometry-Driven Thermodynamics: Shape Effects and Anisotropy in Quantum-Confined Ideal Fermi and Bose Gases.

Rivo Herivola Manjakamanana Ravelonjato[1, a], Ravo Tokiniaina Ranaivoson[1], Raoelina Andriambololona[1], Naivo Rabesiranana[1, 2], Charles Oyverné Randriamaholisoa[3], Wilfrid Chrysante Solofoarisina [1].

[1] Institut National des Sciences et Techniques Nucléaires (INSTN- Madagascar)
*BP 3907 Antananarivo 101, Madagascar, instn@moov.mg*
[2] Department of Physics, Faculty of Sciences – University of Antananarivo
*BP 566 Antananarivo 101, Madagascar*
[3] ESSA, University of Antananarivo
*BP 175 Antananarivo 101, Madagascar*

[a] Corresponding author: *manjakamanana@yahoo.fr*

**Abstract**

This study presents a unified description of the thermodynamics of ideal quantum gases (bosonic and fermionic) under nanoscale confinement, using a Quantum Phase Space (QPS) formalism. We show that the statistical momentum variances $B_{ll}$ provide a unified description of quantum degeneracy: for fermions, they naturally incorporate the Fermi energy, while for bosons, they capture the characteristic energy scale of the condensate. This identification bridges our formalism with established results for harmonically trapped Bose gases, where the oscillator length sets the confinement scale. This key identification allows both Fermi-Dirac and Bose-Einstein statistics to be treated consistently within a single framework. From this foundation, we derive exact analytical expressions for key thermodynamic properties—internal energy, anisotropic pressure tensor, and heat capacity—seamlessly describing the transition from classical to quantum-degenerate regimes. Our results reveal that nanoscale thermodynamics is intrinsically anisotropic: pressure becomes a direction-dependent tensor, with fractional anisotropy reaching unity under extreme confinement, while heat capacity exhibits strong discrete oscillations due to spectral quantization. Notably, pure shape effects, controlled via geometric parameters encoded in $\mathcal{B}_{ll}$, enable the manipulation of phase transitions and thermodynamic responses without altering the size, temperature, or density of the nanoconfined system. Asymptotic analyses confirm rigorous compliance with the third law of thermodynamics and the restoration of classical limits at high temperatures or large sizes. Numerical simulations for confined electron and helium-4 gases demonstrate that quantum effects become significant at experimentally accessible temperatures (from $mK$ to $K$) for confinement scales of $5\ nm - 50\ nm$, with fermionic and bosonic systems exhibiting radically different behaviors under extreme confinement. This work provides a comprehensive theoretical toolkit for predicting and engineering the thermomechanical properties of nanosystems, with direct implications for nanofluidic devices, quantum sensors, and nanostructured materials. Future integration with ab-initio computational tools will enable the validation and application of these predictions to real material systems.

1. Introduction

The thermodynamics of nanoscale systems represents a frontier domain where classical and quantum descriptions converge [1]. While macroscopic thermodynamics relies on the continuum hypothesis and the thermodynamic limit, nanoscale systems exhibit discrete behaviors and strong finite-size effects that challenge established laws [2, 3]. Quantum confinement effects have been experimentally demonstrated in various nanoscale systems, such as $Mo/Si$ multilayer structures used in X-ray and EUV optics, where red shifts of TO phonon modes and blue shifts of plasmon energies were directly linked to the reduction in $Si$ nanocluster size and layer thickness [4] These observations underscore that quantum size effects manifest not only in fundamental electronic and vibrational properties but also in macroscopic optical performance, reinforcing the relevance of confinement-driven thermodynamics in functional nanomaterials. These effects are paradigmatically illustrated in harmonically trapped atomic gases, where Bose-Einstein condensation emerges with spatial and thermodynamic signatures dramatically reshaped by the confinement potential [5]. This interplay is governed by a fundamental competition: when the thermal de Broglie wavelength $\lambda_{th} = \frac{h}{\sqrt{2\pi m k_B T}}$ becomes comparable to the system's characteristic dimensions, quantum statistics and confinement dominate over classical behavior [6]. This regime is not only of fundamental interest but also critical for advancing nanoscale technologies, where the thermal and mechanical properties of confined gases—such as in nanofluidic channels, porous materials, or quantum dot arrays—determine device performance [7, 8, 9]. In particular, quantum size effects have been shown to significantly alter thermal and potential conductivities in such confined geometries [10].

Analytical approaches to confined quantum gases have been developed for regular geometries such as boxes, tubes, and annular containers [11], where exact solutions for grand partition functions reveal boundary and topology effects neglected in the thermodynamic limit. These works highlight that thermodynamic properties become shape-dependent when the thermal wavelength is comparable to the system size—a regime central to nanoscale systems. Beyond geometric confinement, anisotropic hopping parameters in lattice models can also induce profound quantum phase transformations. For instance, in attractively interacting fermionic systems on a square lattice, an anisotropy in next-nearest-neighbor hopping has been shown to stabilize an exotic d-wave Cooper pair Bose metal phase, characterized by a finite momentum Bose surface [12]. This underscores that anisotropy — whether geometric or encoded in effective hopping parameters —serves as a fundamental thermodynamic variable capable of tuning quantum states. In the context of trapped Bose gases, this shape dependence manifests in the anisotropy of collective modes and density profiles, a phenomenon extensively studied in both theory and experiment [5].

Despite significant progress, a unified theoretical framework that seamlessly bridges the quantum and classical regimes while capturing geometric anisotropy remains a challenge. Existing approaches often treat confinement perturbatively or are restricted to symmetric geometries, with [13] being the most notable for its focus on the emergence of a pressure tensor with direction-dependent components—a hallmark of nanoscale thermodynamics.

Recently, a Quantum Phase Space (QPS) formalism has been proposed to address part of these challenges by providing a phase-space—based statistical description of confined quantum systems [14, 15]. Within this framework, the statistical variance of momenta naturally emerges as a key quantity governing thermodynamic behavior under confinement. However, in its previous formulations, this variance was not explicitly connected to the physical manifestations of quantum degeneracy in fermionic and bosonic systems.

In the present work, we show that this same momentum variance provides a unified and explicit description of quantum degeneracy effects when properly interpreted. For fermionic systems, it naturally recovers the Fermi energy per particle, reflecting the Pauli exclusion principle. For bosonic systems, it captures the characteristic energy scale of the condensate below the critical temperature $T_c$ analogous to how the chemical potential in a trapped Bose gas approaches the ground-state energy in the Thomas-Fermi limit [5]. This identification, absent from previous QPS studies, allows both Fermi–Dirac and Bose–Einstein statistics to be treated consistently within a single thermodynamic framework.

Within this extended interpretation of the QPS formalism, we investigate the thermodynamics of ideal quantum gases under arbitrary nanoscale confinement. Our objectives are threefold: (1) to derive exact analytical expressions for key thermodynamic quantities—internal energy, anisotropic pressure, and heat capacity; (2) to systematically quantify the anisotropy induced by asymmetric geometries; and (3) to analyse the asymptotic behaviors at both low and high temperatures, with particular attention to rigorous compliance with the third law of thermodynamic. By achieving these goals, this work provides a comprehensive toolkit for predicting and engineering the thermomechanical properties of nanoscale systems.

## 2. Theoretical Framework and Methodology

### 2.1 Established QPS Formalism

The concept of QPS was introduced, developed, and applied to ideal gas models in refs [14, 15]. Some results from these references are used in the present work.

As shown in ref. [15], within the phase space representation of quantum mechanics, the Hamiltonian operator for a particle $A$ in its mean rest frame is

$$\boldsymbol{h}_A = \sum_{l=1}^{3} (2\boldsymbol{z}_{Al}^\dagger \boldsymbol{z}_{Al} + 1) \frac{\mathcal{B}_{ll}}{m} \qquad (1)$$

where $\boldsymbol{z}_{Al}$ and $\boldsymbol{z}_{Al}^\dagger$ are ladder operators related to the coordinates and momenta operators of the particle, $l$ is the index related to space direction ($l = 1,2,3$), $m$ is the mass of the particle $A$ and $\mathcal{B}_{ll}$ is the ground-state momentum variance in the $l$-$th$ direction of the three dimensional space. This construction shares a conceptual similarity with the description of a particle in a harmonic trap, where the Hamiltonian is diagonalized in terms of creation and annihilation operators, and the characteristic length $a_{h_0 = \hbar/(m\omega)}$ defines the confinement scale [5].

The eigenvalue equations of the hamiltonian operator $\boldsymbol{h}_A$ is

$$\boldsymbol{h}_A |\alpha_A\rangle = \varepsilon_A |\alpha_A\rangle \qquad (2)$$

$|\alpha_A\rangle$ is the eigenstate and $\varepsilon_A$ is the corresponding eigenvalue. The explicit expression of the eigenvalue $\varepsilon_A$ is :

$$\varepsilon_A = \sum_{l=1}^{3} (2n_l + 1) \frac{\mathcal{B}_{ll}}{m} \qquad (3)$$

in which $n_1, n_2, n_3$ are positive integers. Taking account of the relation (3), we can express the eigenstate as $|\alpha_A\rangle$

$$|\alpha_A\rangle = |n_1, n_2, n_3\rangle \qquad (4)$$

The Hamiltonian operator $H_G$ of an ideal gas composed of $N_G$ particles can be deduced using (1) and by summing over all particles.

$$H_G = \sum_{A=1}^{N_G} h_A = \sum_{A=1}^{N_G}\sum_{l=1}^{3}\left(2z_{Al}^{\dagger}z_{Al} + 1\right)\frac{B_{ll}}{m} \qquad (5)$$

An eigenstate $|G\rangle$ of the Hamiltonian operator $H_G$ can be deduced from the one particle energy eigenstates $|\alpha\rangle$ as symmetrized Fock state (for bosons) or antisymmetrized Fock state (for fermions). It can be written in the form

$$|G\rangle = |N(0,0,0); N(1,0,0); N(0,1,0), N(0,0,1); \ldots; N(n_1, n_2, n_3); \ldots\rangle \qquad (6)$$

in which $N_\alpha = N(n_1, n_2, n_3)$ represents the number of particles in a particular state $|\alpha\rangle = |n_1, n_2, n_3\rangle$. We have the eigenvalue equation for the Hamiltonian operator $H_G$ of the gas:

$$H_G|G\rangle = E_G|G\rangle \qquad (7)$$

with the eigenvalue

$$E_G = \sum_\alpha N_\alpha \varepsilon_\alpha = \sum_{n_1}\sum_{n_2}\sum_{n_3} N(n_1, n_2, n_3)\varepsilon(n_1, n_2, n_3) \qquad (8)$$

in which $\varepsilon_\alpha = \varepsilon(n_1, n_2, n_3)$ is the energy value corresponding to a state $|\alpha\rangle = |n_1, n_2, n_3\rangle$

$$\varepsilon_\alpha = \varepsilon(n_1, n_2, n_3) = \sum_{l=1}^{3}(2n_l + 1)\frac{B_{ll}}{m} \qquad (9)$$

## 2.2 Grand Canonical Potential Formalism

The foundation of our approach rests on the grand canonical potential:

$$\Omega = -k_B T \ln \Xi_\alpha \qquad (10)$$

$\Xi_\alpha$: grand canonical partition function

where $\Xi_\alpha$ is associated with quantum states $|\alpha\rangle = |n_1, n_2, n_3\rangle$. For an individual state, partition functions differ according to statistics :

$$\Xi_\alpha = \begin{cases} \sum_{N_\alpha=0}^{\infty} e^{-\beta(\varepsilon_\alpha-\mu)N_\alpha} = \dfrac{1}{1-e^{-\beta(\varepsilon_\alpha-\mu)}} & \text{for bosons} \\ \sum_{N_\alpha=0}^{1} e^{-\beta(\varepsilon_\alpha-\mu)N_\alpha} = 1 + e^{-\beta(\varepsilon_\alpha-\mu)} & \text{for fermions} \end{cases} \quad (11)$$

$\varepsilon_\alpha$ is the energy level corresponding to the state $|\alpha\rangle = |n_1, n_2, n_3\rangle$ and $N_\alpha = N(n_1, n_2, n_3)$ corresponds to the particles number corresponding to the state $|\alpha\rangle$.

### 2.3 Fractional Anisotropy

We employ the Fractional Anisotropy (FA), a standard measure for tensor anisotropy. The expression of (FA) and the idea of measuring it using the eigenvalues of the diffusion tensor first appeared in Refs. [16, 17]. Adapted to our pressure tensor, it is defined as the stress tensor $P_{ii} = \sigma_i$.

$$FA = \sqrt{\dfrac{3}{2}} \cdot \dfrac{\sqrt{\sum_{i=1}^{3}(\sigma_i - \bar{\sigma})^2}}{\sqrt{\sum_{i=1}^{3} \sigma_i^2}} \quad (12)$$

$$Anisotropy = FA \times 100\%$$

### 3. Results and Analysis

### 3.1 Potential Development and Confinement Parameters

Mathematical development of the relations (10) and (11) leads to the expression of the grand canonical potential function:

For both statistics:

$$\Omega = \dfrac{g_s k_B T}{8} \sum_{j=1}^{\infty} \dfrac{s_j}{j} \dfrac{1}{\prod_{l=1}^{3} \sinh\left(j\dfrac{\beta \mathcal{B}_{ll}}{m}\right)} \quad (13)$$

With $s_j = \xi^j$ for bosons and $s_j = (-1)^{j+1}\xi^j$ for fermions, $\xi = e^{\beta\mu}$ is the fugacity.

$n = N/(L_1 L_2 L_3)$: particle density ($m^{-3}$)
$N$: total number of particles
$x_l = j\dfrac{\beta \mathcal{B}_{ll}}{m}$

The statistical variance of momenta $\mathcal{B}_{ll}$ is the central parameter linking geometric confinement to thermodynamic properties. It is rigorously determined by equating the exact discrete quantum mechanical expression for the grand potential (13) with its continuous semi-classical counterpart in two distinct physical limits: the high-temperature/large-volume and the low-temperature/small-volume quantum degenerate regime.

### 3.1.1 Determination of the statistical variance of momenta $\mathcal{B}_{ll}$ in the semi-classical limit: High temperature and Large volume

As shown in Ref. [15], the semi-classical limit is defined by the conditions

$$\begin{cases} \xi = e^{\beta\mu} \ll 1 \\ \beta\dfrac{\mathcal{B}_{ll}}{m} \ll 1 \end{cases} \quad (14)$$

Under these conditions, expanding (13) to leading order yields the discrete representation:

$$\Omega_{disc} \cong -\frac{g_s(m)^3 \xi}{8(\beta)^4 \mathcal{B}_{11}\mathcal{B}_{22}\mathcal{B}_{33}} \quad (15)$$

This is equated with the continuous phase-space integral result:

$$\Omega_{cont} \cong -\frac{g_s V}{\hbar^3}\left(\frac{m}{2\pi\beta}\right)^{3/2} \xi \quad (16)$$

The identification (15)=(16) leads to the purely thermal contribution

$$\mathcal{B}_{ll} = \frac{\hbar}{2L_l}\sqrt{2\pi m k_B T} \quad (17)$$

### 3.1.2 Determination of the statistical variance or momenta in the quantum degenerate limit: low temperature and small system size.

While the QPS formalism was introduced in Ref. [15], the present work constitutes a substantial advance by extending it to quantum-degenerate Fermi and Bose gases. In the degenerate regime ( $T \to 0$ fermions and $T \to T_c$ for bosons), where the semiclassical conditions underlying Ref. [15] break down, the momentum variances $\mathcal{B}_{ll}$ are no longer fixed by semi-classical arguments but are instead determined by matching the particle density $n = N/V$ obtained from the exact discrete expression.

$$n_{disc} = -\frac{1}{V}\left(\frac{\partial\Omega}{\partial\mu}\right)_{T,V} = \frac{1}{8V}\sum_{j=1}^{\infty}\frac{js_j}{\prod_{l=1}^{3}\sinh(jx_l)} \quad (18)$$

With the standard result from the continuous theory of quantum gases,

$$n_{cont} = \frac{g_s}{\lambda_{th}^3}f_{3/2}(\xi) \quad (19)$$

Analyzing this equality in the respective degenerate limits yields the quantum-dominated contributions. The details of the derivation are provided in the Appendix.

For fermions at $T \to 0$: the chemical potential approaches the Fermi energy, $\mu \to \epsilon_F = \frac{\hbar^2}{2m}\left(\frac{6\pi^2}{g_s}\right)^{2/3}$. The matching procedure gives:

$$\mathcal{B}_{ll}^{(quant,Fermions)} = \frac{\hbar}{2L_l}\sqrt{\pi\hbar^2\left(6\pi^2\frac{n}{g_s}\right)^{\frac{2}{3}}} \qquad (20)$$

For bosons at $T \to T_c$: the condition for condensation gives $n = \frac{g_s}{\lambda_{th}^3}\zeta(3/2)$. The matching procedure yields:

$$\mathcal{B}_{ll}^{(quant,Bosons)} = \frac{\hbar}{2L_l}\sqrt{\frac{4\pi\hbar^2}{m}\left(\frac{n}{g_s\zeta\left(\frac{3}{2}\right)}\right)^{\frac{2}{3}}} \qquad (21)$$

### 3.1.3 Unified interpolating form for all regimes

The results from the two distinct physical limits motivate a unified, interpolating form for $\mathcal{B}_{ll}$ that captures the smooth transition from the classical to the quantum regime. The structure $\sqrt{A+B}$ represents the quadrature addition of momentum uncertainties from independent thermal ($A$) and quantum statistical ($B$) sources, which is the neutral form for a variance.

For fermions

$$\mathcal{B}_{ll} = \frac{\hbar}{2L_l}\sqrt{2\pi mk_BT + \pi\hbar^2\left(6\pi^2\frac{n}{g_s}\right)^{\frac{2}{3}}} \qquad (22)$$

For bosons

$$\mathcal{B}_{ll} = \frac{\hbar}{2L_l}\sqrt{2\pi mk_BT + \frac{4\pi\hbar^2}{m}\left(\frac{n}{g_s\zeta\left(\frac{3}{2}\right)}\right)^{\frac{2}{3}}} \qquad (23)$$

$n$ corresponds to the density of particles, $n = N/V$.

These expressions for $\mathcal{B}_{ll}$, derived from first principles within the QPS formalism, provide a fundamental connection between geometric confinement and quantum degeneracy. Interestingly, the structure of these expressions reveals a deep physical insight: the second term under the square root corresponds to the Fermi energy per particle for fermions and to the characteristic energy scale of the condensate for bosons. This establishes a direct conceptual bridge with the energy scales that govern the thermodynamics of trapped quantum gases, such as the chemical potential in the Thomas-Fermi limit for Bose-Einstein condensates [5], while arising from an entirely different theoretical framework. Our results thus offer an independent, phase-space-based confirmation of these fundamental energy scales, now explicitly linked to confinement parameters through $\mathcal{B}_{ll}$.

## 3.2 Internal Energy and Asymptotic Behaviors

### 3.2.1 Exact expression of internal energy

The internal energy is derived from the relation $U = \left(\frac{\partial(\beta\Omega)}{\partial\beta}\right)_{V,\mu}$

$$U = \frac{g_S k_B T}{8} \sum_{j=1}^{+\infty} \frac{s_j}{\prod_{k=1}^{3} \sinh(jx_k)} \left[\frac{\mu}{k_B T} - j \sum_{l=1}^{3} x_l \coth(jx_l)\right] \quad (24)$$

### 3.2.2 Asymptotic behavior of internal energy

#### 3.2.2.1 *Asymptotic behavior according to the temperature*

➢ Low temperature
  For fermions $T \ll T_F$ ($T \to 0$):
$$U_{fermions} \approx \frac{3}{5} N\epsilon_F + \frac{\pi^2}{4} Nk_B \frac{T^2}{T_F} \quad (25)$$
  For bosons $T \ll T_c$
$$U_{bosons} \approx \frac{2\pi^2 \zeta\left(\frac{5}{2}\right)}{\zeta(3/2)} Nk_B T \left(\frac{T}{T_c}\right)^{\frac{3}{2}} \quad (26)$$

For fermions, the internal energy approaches a finite zero-point value $\frac{3}{5}N\epsilon_F$, representing the irreducible kinetic energy of the filled Fermi sea—a direct manifestation of the Pauli exclusion principle. The quadratic temperature correction arises from thermal excitations near the Fermi surface. In contrast, bosonic systems exhibit a power-law approach to zero energy as $T \to 0$, reflecting the collective condensation into the ground state with minimal residual energy.

For confined bosons, the low temperature behavior modifies to $U_{confined}^{bosons}(T \to 0) \approx ANk_B T \left(\frac{T}{T_c}\right)^2$. Both satisfy the third law of thermodynamics with $U(T = 0)$ representing the absolute minimum.

➢ High temperature $T \to \infty$
Demonstrate universal classical recovery:

$$U(T \to \infty) = \frac{3}{2} Nk_B T \left[1 - \frac{\pi^3}{2}\left(\frac{\hbar^2}{mk_B TL^2}\right)^{\frac{3}{2}} + \mathcal{O}\left(\frac{\hbar^4}{m^2 k_B^2 T^2 L^4}\right)\right] \quad (27)$$

This high-temperature asymptotic form is identical for both fermions and bosons, demonstrating that quantum statistical differences become negligible when thermal energy dominates all quantum scales.

$$\lim_{T \to \infty} U = \frac{3}{2} N k_B T \qquad (28)$$

Equation (5) represents the equipartition theorem's restoration-each translational degree of freedom contributes $\frac{1}{2} k_B T$, independent of quantum statistics. The quantum corrections in equation (27), appearing as powers of $\frac{\hbar^2}{m k_B T L^2}$, systematically vanish at high temperature, illustrating the correspondence principle in action.

*3.2.2.2 Asymptotic behavior according to the system sizes*

$$\lim_{L \to \infty} U = \frac{3}{2} N k_B T \qquad (29)$$

$$\left. \begin{array}{l} U_{fermions}(L \to 0) \propto L^{-5} e^{-\frac{C}{L^2}} \to 0 \\ U_{bosons}(L \to 0) \propto L^{-4} e^{-\frac{C}{L^2}} \to 0 \end{array} \right\} \qquad (30)$$

Equation (29) confirms extensive thermodynamics in macroscopic limit, while Equations (30) reveal dramatically different approaches to zero energy under extreme confinement-fermionic energy decays with power-law divergence tempered by exponential suppression, whereas bosonic energy vanishes rapidly with $L^4$ scaling. The exponential suppression factor $e^{-\frac{C}{L^2}}$ originates directly from the behavior of $\sinh(j x_l)$ in (13) when $x_l = \beta \mathcal{B}_{ll}/m \propto 1/L \to \infty$, as derived from the unified expression (22) and (23).

## 3.3 Anisotropic Pressure and Geometric Dependence

### 3.3.1 Exact expression of anisotropic pressure

From the relation $P = -\frac{\partial \Omega}{\partial V}$, but accounting for anisotropy, we derive the pressure tensor:

$$P_{kk} = \frac{g_S k_B T}{8V} \sum_{j=1}^{+\infty} \frac{s_j}{\prod_{k=1}^{3} \sinh(j x_k)} [j x_l \coth(j x_l)] \qquad (31)$$

$P_{kk}$: directional pressure component
$V$: system volume

### 3.3.2 Asymptotic behaviors of pressure

*3.3.2.1 Asymptotic behavior according to the temperature*

➢ Low temperature

For fermions $T \ll T_F$ ($T \to 0$)

$$P_{kk}^{fermions} \approx \frac{8}{15} n \epsilon_F \left[ 1 + \frac{5\pi^2}{12} \left( \frac{T}{T_F} \right)^2 \right] \qquad (32)$$

For bosons $T \ll T_c$

$$P_{kk}^{bosons} \approx \frac{\pi^2}{45} nk_B T \left(\frac{T}{T_c}\right)^{\frac{3}{2}} \tag{33}$$

> High temperature $T \to \infty$
> ($T \gg T_F$) and ($T \gg T_c$)

$$P_{kk}(T \to \infty) = \frac{Nk_B T}{V}\left[1 - \frac{\pi^3}{4}\left(\frac{\hbar^2}{mk_B TL^2}\right)^{\frac{3}{2}} + \mathcal{O}\left(\frac{\hbar^4}{m^2 k_B^2 T^2 L^4}\right)\right] \tag{34}$$

$$\lim_{T \to \infty} P_{kk} = \frac{Nk_B T}{V} \text{ for all } k = 1, 2, 3 \tag{35}$$

Equation (35) represents the restoration of the ideal gas law and complete isotropy-thermal motion averages out directional preferences imposed by confinement geometry. The fractional anisotropy $FA(T \to \infty) = 0$ quantitatively confirms this isotropic recovery.

*3.3.2.2 Asymptotic behavior according to the system sizes*

$$\lim_{L \to \infty} P_{kk} = \frac{Nk_B T}{V}, FA(L \to \infty) = 0 \tag{36}$$

$$P_{kk}^{fermions}(L \to 0) \propto L^{-5}e^{-C/L^2} \to 0 \tag{37}$$

$$P_{kk}^{bosons}(L \to 0) \propto L^{-4}e^{-C/L^2} \to 0, FA(L \to 0) = 1 \tag{38}$$

Equation (35) confirms pressure becomes extensive and isotropic in the thermodynamic limit. Equations (36) and (37) show both pressures approach zero under extreme confinement, but with different functional forms and maximum anisotropy ($FA \to 1$) reflecting dominant boundary effects. The exponential decay with $L$, as in (30), is a direct consequence of the unified $\mathcal{B}_{ll}$ expressions (16) and (17) in the limit $L \to 0$.

## 3.4 Heat Capacity and Third Law Compliance

### 3.4.1 Exact expression of heat capacity

From $C_v = \left(\frac{\partial U}{\partial T}\right)_{N,V}$, after meticulous calculation:

$$C_v = \frac{g_s k_B}{16}\left[\sum_{j=1}^{\infty} \frac{s_j}{D^{(j)}} A_j + \left(\frac{\mu}{T} + \frac{d\mu}{dT}\right)\sum_{j=1}^{\infty} \frac{s_j}{D^{(j)}} B_j\right] \tag{39}$$

With

- $A_j = RN_0^{(j)}\left[1 - \frac{(R-1)(2-R)}{2k_B T}\right] - \frac{R(3-R)}{2k_B T}\sum_{l=1}^{3} jx_l\left[\coth(jx_l) - \frac{jx_l}{(\sinh(jx_l))^2}\right] + \frac{R(3-R)T}{2}\left(\frac{N_0^{(j)}}{j}\right)^2$

- $D^{(j)} = \prod_{k=1}^{3} \sinh(jx_k)$ ; $N_0^{(j)} = \sum_{l=1}^{3} x_k \coth(jx_k)$ ; $B_j = \frac{RT}{k_B} j N_0^{(j)}$

- $\frac{d\mu}{dT} = -k_B T \frac{\sum_{j=1}^{\infty} \frac{js_j}{D^{(j)}} \left[\frac{3-R}{2T} N_0^{(j)} + \frac{j\mu}{k_B T}\right]}{\sum_{j=1}^{\infty} \frac{j^2 s_j}{D^{(j)}}} - \frac{\mu}{T}$

The factor $R = R(T)$ appears here as it is part of the interpolating structure developed to simplify the exact but complex expression for $C_v$.

$$R(T) = \begin{cases} R_F(T) = 1 + \dfrac{T_F/T}{1 + T_F/T} \\ R_B(T) = 1 + \dfrac{T_c/T}{1 + T_c/T} \end{cases} \tag{40}$$

The quantumness parameter $R(T)$ interpolates smoothly between classical ($R = 1$) and quantum degenerate ($R = 2$) regimes.

$T_F = \frac{\hbar}{2m\pi k_B} \left[6\pi^2 \frac{N}{L^3}\right]^{2/3}$ and $T_c(L) = \frac{\hbar}{2mk_B} \left[\frac{1}{\zeta(\frac{3}{2})} \cdot \frac{N}{L^3}\right]^{2/3}$ refers respectively to the Fermi temperature and to the critical temperature for Bosons gases.

3.4.2    Asymptotic behavior of heat capacity

*3.4.2.1   Asymptotic behavior according to temperature*

➢ Low temperatures
- For fermions $T \ll T_F$ ($T \to 0$)

$$C_v^{fermions} \approx \frac{\pi^2}{2} N k_B \frac{T}{T_F} \to 0 \tag{41}$$

This linear dependence is the distinctive signature of a degenerate Fermi gas. It reflects the fact that only electrons within an energy band of width $k_B T$ around the Fermi level can participate in thermal excitation. The slope contains valuable information about the density of states at the Fermi level. The limit $\lim_{T \to 0} C_v = 0$ rigourously satisfies the third law of thermodynamics.

- For bosons $T \ll T_c$ ($T \to 0$)

$$C_v^{bosons} \approx 1.925 N k_B \left(\frac{T}{T_c}\right)^2 \to 0 \tag{42}$$

The quadratic dependence at low temperature differs from the $T^{3/2}$ behavior of macroscopic bosonic systems, thus revealing the specific effects of confinement. As with fermions, the nullity at zero temperature is guaranteed.

➢ High temperature ($T \to \infty$)
- For both statistics $T \gg T_F$ and $T \gg T_c$

$$C_v(T \to \infty) = \frac{3}{2}Nk_B \left[1 - \frac{3\pi^3}{4}\left(\frac{\hbar^2}{mk_BTL^2}\right)^{\frac{3}{2}} + \mathcal{O}\left(\frac{\hbar^4}{m^2k_B^2T^2L^4}\right)\right] \quad (43)$$

$$\lim_{T \to \infty} C_v = \frac{3}{2}Nk_B \quad (44)$$

At high temperature, the Maxwell-Boltzmann expression of heat capacity is recovered for both statistics.

*3.4.2.2 Asymptotic behavior according to the system sizes*

➢ *Macroscopic limit ($L \to \infty$)*

$$\lim_{L \to \infty} C_v \approx \frac{3}{2}Nk_B \quad (45)$$

This convergence toward the classical Dulong-Petit limit illustrates the correspondence principle: when confinement becomes negligible, quantum properties fade in favor of classical behavior. Physically, the characteristic energy scale $T_F$ or $T_c$ becomes infinitely small compared to thermal agitation, making quantum indistinguishability effects negligible. The value $\frac{3}{2}Nk_B$ corresponds exactly to the equipartition theorem for three translational degrees of freedom.

➢ *Extreme confinement limit (L very small)*

$$\lim_{L \to 0} C_v^{fermions} \approx \frac{\pi^2 C_F}{1.2T} Nk_B L^{-2} \to \infty \; (\propto L^{-2}) \quad (46)$$

Since $T_F(L) = C_F L^{-2}$

For extreme confinement, fermionic heat capacity diverges as $L^{-2}$ due to enhanced density of states and quantum degeneracy pressure.

$$\lim_{L \to 0} C_v^{bosons} \approx \frac{3.85(1+\alpha)T^{\frac{3}{2}}}{C_c^{\frac{3}{2}}} Nk_B L^3 \to 0 \; (\propto L^3) \quad (47)$$

Since $T_c(L) = C_c L^{-2}$

For extreme confinement, bosonic heat capacity vanishes as $L^3$ due to perfect condensation in the ground state with minimal thermal fluctuations.

3.4.3  Interpolating function of heat capacity

The interpolating function for $C_v$ is obtained by expanding the exact expression derived from the phase space formalism.

For fermions:

$$C_v^{fermions}(T,L) = Nk_B \left[\frac{\pi^2}{2}\frac{T}{T_F}\frac{R_F(T)}{1.2\left(\frac{T}{T_F}\right)^2} + \frac{3}{2}\frac{1}{1+1.5\left(\frac{T_F}{T}\right)^2}\right] \tag{48}$$

Maximum : $C_v^{max}$ 3.22 $Nk_B$ à $T = 0.34T_F$

This expression skillfully combines the low-temperature linear regime and the classical saturation at high temperature. The presence of the factor $R_F$ incorporates quantum corrections that becomes significant when temperature approaches $T_F$.

For bosons:

$$C_v^{bosons} = \begin{cases} 1.925Nk_B \left(\frac{T}{T_c}\right)^{\frac{3}{2}}\left[1+\alpha\left(\frac{T_c-T}{T_c+T}\right)\right]R_B(T) & T<T_c \\ 1.925Nk_B\left[\frac{1.5}{1.925}+0.221\left(\frac{T_c}{T}\right)^{\frac{3}{2}}\right]R_B(T) & T>T_c \end{cases} \tag{49}$$

$\alpha = 0.3$

Maximum : $C_v^{max} \approx 2.89\, Nk_B$ à $T = T_c$

This piecewise form captures the essence of the Bose-Einstein condensation phenomenon, with different power dependencies on either side of the critical temperature. The correction term $\alpha$ adjusts the curvature near $T_c$.

### 3.4.4 Numerical simulation and data

| Fermions (electrons) | Bosons (Helium-4) |
|---|---|
| - $N$ : 1000 electrons | - N = 1000 atoms |
| - $L_1 = L_2 = L_3 = 5\, nm - 50\, nm$ | - $L_1 = L_2 = L_3 = 5\, nm - 50\, nm$ |
| - $m = 9.11 \times 10^{-31} kg$ | - $m = 6.64 \times 10^{-2} kg$ |
| - $g_s = 2\, (spin\frac{1}{2})$ | - $g_s = 1$ |
| Density $n = 10^{29}\, particles/m^3$ ||

The following graphics show the effects of confinement on fermion and bose gases for dimensions ranging from 50 nm to 5.

Figure 1 (Fermions Heat Capacity for different confinement sizes) illustrates several key features: (1) systematic peak shifts following $T_F \propto L^{-2}$ scalling-from $10^3 K$ for $L = 50\, nm$ to $\sim 10^5 K$ for $L = 5\, nm$; (2) Broad, symmetric peaks characteristic of fermionic crossover (not phase transition) with maxima at $\sim 0.34 T_F$ and heights of $3.22Nk_B$; (3) Linear low-temperature behavior visible as straight segments on the log-log plot; (4) Universal convergence to the classical limits $\frac{3}{2}Nk_B$ at high temperature.

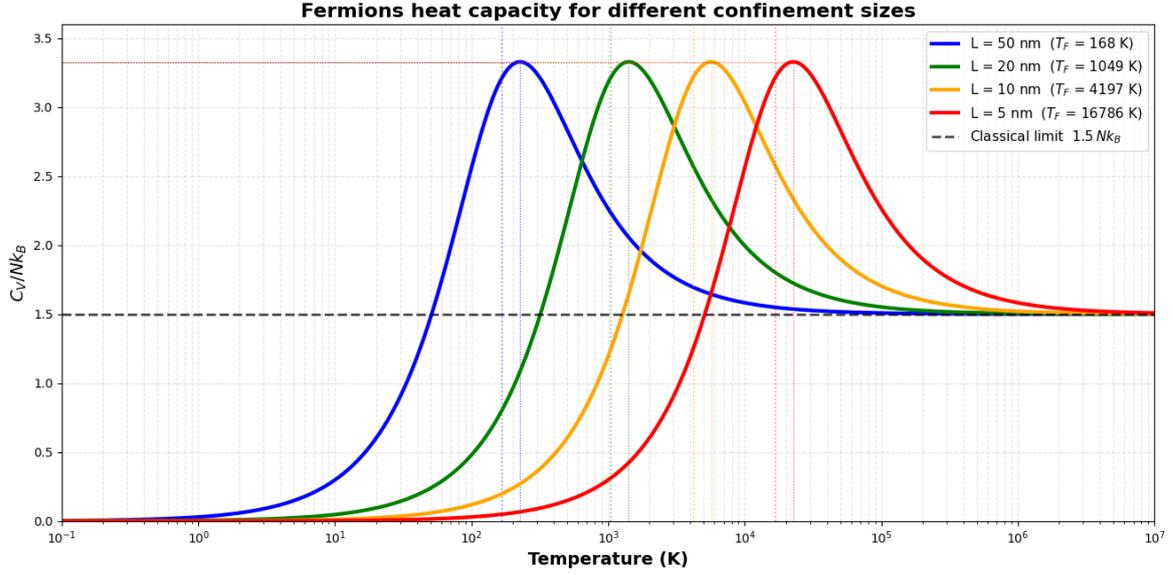

Figure 1: Effects of confinement on fermion gases for dimensions ranging from 50 nm to 5 nm.

These observations confirm that fermionic quantum effects becomes significant at experimentally accessible temperature ($10^3 - 10^5 K$) for nanoscale confinement, with peak positions tunable via simple sizes control.

Figure 2 (Bosons Heat Capacity for Different Confinement Sizes) reveals contrasting behavior: (1) Sharp, asymmetric peaks located exactly at $T_c$, with position scaling as $T_c \propto L^{-2}$—From $\sim 10^4 K$ for $L = 50\ nm$ to $\sim 10^{-2} K$ for $L = 5\ nm$; (2) extremely low temperature scales requiring advanced cryogenic techniques; (3) Peak heights of approximately $2.88 N k_B$, exceeding the classical limit by a factor 1.92; (4) Rapid decay above $T_c$ following a $T^{-3/2}$ power law.

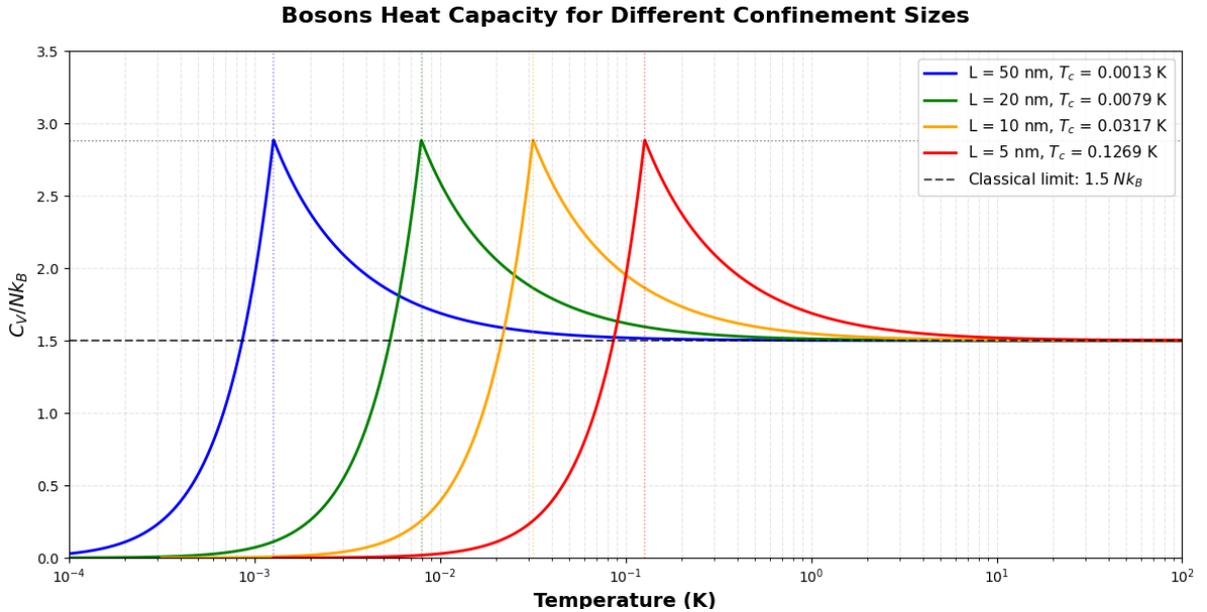

Figure 2: Effects of confinement on Boson gases for dimensions ranging from 50 nm to 5 nm.

The peak sharpness signals a genuine second-order phase transition (Bose-Einstein condensation), while the asymmetry reflects different critical exponents.

The comparative analysis reveals fundamental statistical fingerprints: fermions exhibit broad maxima from Pauli blocking statistics, while bosons show sharp peaks from condensation transitions; fermionic low-temperature behavior is linear ($C_v \propto T$), bosonic is quadratic ($C_v \propto T^2$); identical confinement produces dramatically different size scaling-fermionic response enhances ($C_v \uparrow$ as $L \downarrow$). Both systems share universal features: $T_{char} \propto L^{-2}$ scaling, convergence to classical limits at high temperature, and third law compliance.

4. Discussion

This study presents a unified formalism based on Quantum Phase Space (QPS) to describe the thermodynamic properties of ideal quantum gases confined at the nanoscale. Our approach provides exact analytical expressions for internal energy, anisotropic pressure, and heat capacity, explicitly linking these quantities to geometric confinement parameters. via the statistical variances of momenta $B_{ll}$. These variances, defined as $\mathcal{B}_{ll} = b(T, L_l)/L_l$ for a parallelepipedic box, incorporate both classical thermal contributions and quantum degeneracy effects, enabling a continuous description from the classical to the quantum regime.

Our results reveal that thermodynamics at the nanoscale is intrinsically anisotropic. Pressure does not reduce to an isotropic scalar but forms a tensor whose components $P_{kk}$ depend on direction, a feature directly linked to confinement geometry via $L_l$ and amplified by quantum statistics via the parameter $R(\beta)$. The Fractional Anisotropy ($FA$) index quantifies this directional dependence, reaching its maximum ($FA \to 1$) under extreme confinement, where boundary effects dominate.

Our work is situated within the broader field of thermodynamics beyond the macroscopic limit. It is instructive to compare it with the foundational framework of nanothermodynamics proposed by Hill [18] and with exact statistical mechanical treatments of confined ideal gases [11]. While Hill introduces a subdivision potential $L$ to describe an ensemble of independent small systems, and Dai & Xie employ Euler–Maclaurin summation to extract geometry-dependent terms from the grand potential, our model addresses the geometric and quantum non-ideality within a single system. Our parameters $\mathcal{B}_{ll}$, related to momentum variance, play a role analogous to an "effective geometric potential," encapsulating the effect of confinement on the energy spectrum. A fundamental difference lies in the control variables: Hill's formalism emphasizes the dependence of a small system's properties on environmental variables $(T, P, \mu)$, while our study isolates the effect of intrinsic system variables – its size $L$ and shape via $\mathcal{B}_{ll}$ – at fixed density and temperature. These two perspectives, one statistical-ensemble-based and the other mechanical-microscopic, converge to emphasize that nanoscale thermodynamics cannot be reduced to a simple scaling of macroscopic laws. This reflection connects with fundamental questions raised by quantum thermodynamics of small systems [19].

The analysis of asymptotic behaviors under extreme confinement ($L \to 0$) finds resonance in the rigorous framework of confined fluid thermodynamics. Recent work [20] demonstrated that classical thermodynamics, with careful handling of singularities, remains valid for describing the $3D \to 2D$ transition as the width of a slit pore tends to zero. The renormalization procedure they implement to subtract divergences in $1/L$ and $\ln L$ is conceptually analogous to our approach. Although our study deals with quantum gases and theirs with a classical hard-sphere fluid, the philosophy is common: the effective thermodynamic quantities (like our anisotropic pressure $P_{kk}$ or heat capacity $C_v$ must be understood as well-defined limits after subtracting singular contributions induced by confinement geometry.

Our study, centered on the influence of geometric parameters via $\mathcal{B}_{ll}$, aligns with efforts to exploit pure shape effects in confined systems [13]. The physical relevance of such geometry-driven control is underscored by experimental observations across diverse nanoscale platforms: in graphene/WSe$_2$ heterostructures, interfacial phase transitions and domain wall movements actively reshape electrostatic potentials and quasibound states [21], while in Mo/Si multilayer mirrors, phonon and electron confinement manifest as red-shifted Raman modes and blue-shifted plasmon energies with decreasing layer thickness [4]. These results collectively validate that quantum confinement is not merely a theoretical construct but a measurable, tunable phenomenon in functional nanomaterials, reinforcing the role of parameters like $\mathcal{B}_{ll}$ that encode geometric control. Interestingly, comparable anisotropic effects are observed in correlated electron systems, where anisotropic next-nearest-neighbor hopping can promote a d-wave Bose metal phase, altering pair symmetry and ground-state order [12]. Although, the microscopic mechanism differ—Confinement via hard walls versus hopping anisotropy in lattice models —both cases highlight anisotropy as a controlling parameter for thermodynamic and quantum phase behavior. Experimental work with ultracold quantum gases has shown that a size-invariant geometric transformation can induce and control Bose-Einstein condensation without changing temperature, density, or size parameters [22]. This shape-induced condensation is accompanied by pronounced thermodynamic signatures, underscores the central role of geometry as a thermodynamic variable, a principle that our formalism captures through the parameters $\mathcal{B}_{ll}$ [5]. These results support our approach, which attributes to the parameters $\mathcal{B}_{ll}$ the role of geometric information carriers. Furthermore, prior theoretical work on quantum shape effects predicted unusual thermodynamic behaviors, such as the simultaneous decrease of entropy and free energy during a shape transformation, and the emergence of torque due to anisotropic pressure [10, 13, 22, 23, 24].

These effects find an echo in condensed matter systems. A pioneering study on CoPt alloy nanoparticles [24] demonstrated that the order-disorder transition temperature is governed not only by size but crucially by the 3D shape and morphology of the nanoparticle, being dictated by its smallest characteristic dimension. This observation provides a material validation of the principle that geometry is a fundamental thermodynamic parameter, reinforcing our conclusion that the parameters $\mathcal{B}_{ll}$, are adequate to describe such effects.

Our detailed analysis of the heat capacity of confined Fermi gases reveals a discrete and oscillatory nature as a function of particle number or size, in the regime of strong degeneracy and confinement. This behavior, marked by staircases in chemical potential and sharp peaks in $C_v$, is a direct manifestation of the discretization of the energy spectrum and the Fermi surface structure [3]. It confirms the necessity of an exact treatment beyond the continuous approximation.

The asymptotic behaviors rigorously respect fundamental laws. At high temperatures or large sizes, classical limits (equipartition) are recovered. At low temperature, $C_v \to 0$ for both statistics, ensuring compliance with the third law. Our equilibrium approach is complementary to advances in nonequilibrium thermodynamics of small systems, where fluctuation theorems connect single-trajectory work measurements to equilibrium free energy differences [25], a connection potentially exploitable for experimental validation of our predictions.

The comparison between fermions and bosons reveals distinct statistical fingerprints. For fermions, $C_v$ exhibits a broad, symmetric peak around $\sim 0.34\, T_F$, reflecting a crossover. For bosons, a sharp, asymmetric peak appears at $T_c$, signaling a second-order phase transition. These peaks are modulated by system size ($T_F, T_c \propto L^{-2}$), offering a means of control via geometry.

Numerical simulations confirm that quantum effects become significant at accessible temperatures *(from mK to K)* for confinement sizes of 5 to 50 nm. The divergence of $C_v$ for fermions and its vanishing for bosons under extreme confinement ($L \to 0$) highlight the opposite responses of the two statistics to geometric constraints.

## 5. Conclusion

The Quantum Phase Space formalism developed here offers a unified and fundamental description of the thermodynamic properties of ideal quantum gases under nanoscale confinement. We have demonstrated that the parameters $\mathcal{B}_{ll}$ serve as a unique link between confinement geometry, quantum statistics, and macroscopic observables: internal energy, anisotropic pressure, and heat capacity.

The main findings establish that:

- Nanoscale thermodynamics is intrinsically anisotropic, with a tensorial pressure and maximum anisotropy under extreme confinement.
- Pure shape effects, via geometric parameters encoded in $\mathcal{B}_{ll}$, constitute a degree of freedom for manipulating phase transitions and thermodynamic responses, which is corroborated by experiments on quantum gases [22], nanoparticles [24], and theoretical predictions [23].
- The formalism captures the transition between classical and quantum regimes and reveals the discrete and oscillatory nature of the properties of strongly confined Fermi gases [3].
- The response to extreme confinement ($L \to 0$) is radically different according to statistics, reflecting the opposition between the Pauli exclusion principle and Bose-Einstein condensation.

This work provides a theoretical toolkit for modeling systems where quantum confinement effects dominate (nanochannels, optical traps, nanostructured materials) and complements studies on confinement-induced modifications of transport coefficients [10]. Our findings, together with recent studies on hopping-anisotropè-induced Bose metal phases in lattice fermion systems [12], demonstate that anisotropy—wether geometric or interaction-derived—is a powerful design variable for controlling quantum states and thermodynamic properties at the nanoscale. It opens prospects for the design of quantum sensors exploiting thermodynamic sensitivity to shape and size, and for the engineering of materials with controlled thermomechanical properties. The practical relevance of geometry-driven confinement is illustrated by experimental systems where quantum states are actively tuned via interfacial engineering—such as STM-induced phase transitions in van der Waals heterostructures [21]—or where optical performance in multilayer mirrors is enhanced by tailoring quantum size effects [4]. These examples demonstrate that quantum confinement is not merely a fundamental curiosity but a design variable for reconfigurable quantum devices and optimized functional materials.

More broadly, this study contributes to the effort of building a predictive and unified nanoscopic thermodynamics, engaging in dialogue with foundational approaches to nanothermodynamics [18], treatments of confinement limits [20], quantum thermodynamics [19], and nonequilibrium methods [25]. Like the bond-energy model for solids [26], it participates in a common framework showing that thermodynamic properties at the nanoscale are governed by a fundamental perturbation of the coordination environment or energy spectrum due to geometric constraints.

Our approach thus complements earlier exact methods [11] by providing a unified phase-space framework for anisotropic, quantum-degenerate confined gases, opening new pathways for engineering nanoscale thermodynamics response. To concretize these perspectives, future work will integrate the QPS formalism with ab-initio codes such as Quantum ESPRESSO and the EPW/elphbolt modules. This synergy will allow validation of predictions on real material systems (electron gases in semiconductor nanostructures, 2D heterostructures) and quantification of the impact of confinement effects on electronic and thermal transport in nanostructured materials. This convergence between the fundamental theory of confined quantum gases and condensed matter simulation tools represents a promising step toward the rational design of quantum materials and devices with optimized thermal and electronic functionalities.

# Appendix: Determination of the Statistical Variance of Momenta in the Quantum Degenerate Limit

## A.1. Context and Approach

In the quantum degenerate limit (low temperature for fermions, temperature near the critical condensation point for bosons), the semi-classical conditions $\xi = e^{\beta\mu} \ll 1$ and $\beta\mathcal{B}_{ll}/m \ll 1$ break down. Determining the confinement parameter $\mathcal{B}_{ll}$ then requires a different approach from the high-temperature limit. Instead of directly equating the expressions for the grand potential, we equate the expressions for the particle density $n = N/V$ obtained from the discrete (exact quantum) and continuous (standard quantum gas theory) representations of the system. Analyzing this equality in the respective degenerate limits allows us to isolate the purely quantum contribution to $\mathcal{B}_{ll}$.

## A.2. Density Expression in the QPS Formalism

Starting from the exact expression for the grand potential $\Omega$ given by equation (13) in the main text, the particle density is obtained via the thermodynamic derivative:

$$n_{disc} = -\frac{1}{V}\left(\frac{\partial\Omega}{\partial\mu}\right)_{T,V}$$

Performing this derivation yields the following discrete expression:

$$n_{disc} = \frac{1}{8V}\sum_{j=1}^{\infty}\frac{js_j}{\prod_{l=1}^{3}\sinh(jx_l)} \qquad (A.1)$$

Where $s_j = \xi^j$ for bosons and $s_j = (-1)^{j+1}\xi^j$ for fermions, $\xi = e^{\beta\mu}$ is the fugacity and $x_l = j\frac{\beta\mathcal{B}_{ll}}{m}$

## A.3. Density Expression in the Continuous Theory of Ideal Quantum Gases

In parallel, the standard theory of ideal quantum gases provides a well-known expression relating density, fugacity $\xi = e^{\beta\mu}$, and temperature via the thermal de Broglie wavelength $\lambda_{th}$.

$$n_{cont} = \frac{g_s}{\lambda_{th}^3}f_{3/2}(\xi) \qquad (A.2)$$

Where $g_s$ is the spin degeneracy and $f_{3/2}(\xi)$ denotes the Fermi-Dirac integral for fermions or the Bose-Einstein integral for bosons, of order of $3/2$

## A.4. Fermion Limit at Zero Temperature ($T \to 0$)

For a Fermi gas at zero temperature, the Pauli exclusion principle fills all energy states up to the Fermi energy $\epsilon_F$. The chemical potential approaches this energy, $\mu \to \epsilon_F$ which is given by:

$$\epsilon_F = \frac{\hbar^2}{2m}\left(\frac{6\pi^2}{g_s}\right)^{2/3} \quad (A.3)$$

In this limit, the function $f_{3/2}(\xi)$ behaves asymptotically as $f_{3/2}(e^{\beta\epsilon_F}) \to (2/3\sqrt{\pi})(\beta\epsilon_F)^{2/3}$. Thus, the continuous density expression becomes:

$$n_{cont}(T \to 0) \approx \frac{g_s}{\lambda_{th}^3} \cdot \frac{2}{3\sqrt{\pi}}(\beta\epsilon_F)^{2/3} \quad (A.4)$$

Substituting the expression for $\lambda_{th}$ and $\epsilon_F$ confirms that this returns $n$. Equating this asymptotic form of $n_{cont}$ with the dominant behavior of the discrete expression $n_{disc}$ in the same limit allows us to solve the equation for $\mathcal{B}_{ll}$. Assuming form $\mathcal{B}_{ll} = \frac{\hbar}{2L_l}F$, where $F$ is a function to be determined, the calculation leads to:

$$F^2 = \pi\hbar^2\left(\frac{6\pi^2 n}{g_s}\right)^{2/3} \quad (A.5)$$

From this, we deduce the quantum contribution for fermions:

$$\mathcal{B}_{ll}^{(quant, Fermions)} = \frac{\hbar}{2L_l}\sqrt{\pi\hbar^2\left(6\pi^2 \frac{n}{g_s}\right)^{\frac{2}{3}}} \quad (A.6)$$

This expression is proportional to the square root of the Fermi energy per particle, thus capturing the characteristic energy scale of the fully degenerate Fermi gas.

### A.5. Boson Limit at the critical temperature ($T \to T_c$)

For a Bose gas, the key phenomenon is Bose-Einstein condensation. The critical temperature $T_c$ is defined by the condition where the chemical potential approaches zero from below ($\mu \to 0^-$) while the function $f_{3/2}(\xi)$ reaches its maximum finite value, given by the Rieman zeta function:

$f_{3/2}(1) = \zeta(3/2) \approx 2.612$. At this temperature, the gas density is therefore given by:

$$n_{cont} = \frac{g_s}{\lambda_{th}^3(T_c)}\zeta_{3/2} \quad (A.7)$$

This relation implicitly defines $T_c$ as a function of $n$. To determine $B_{ll}$ in this limit, we proceed analogously to the fermionic case. We analyze the behavior of the equality $n_{disc} = n_{cont}$ as $T$ approaches $T_c$ and $\xi \to 1$. In this limit, the sum in $n_{disc}$ is dominated by the behavior of the $sinh$ functions for small values of their arguments (since $T_c$ is not an extremely low temperature for a macroscopic system). Inverting the resulting series, after substituting relation (A.7), allows us to isolate the dependence of $n$ and $T_c$ in the expression for $B_{ll}$.

The result of this procedure is :

$$\mathcal{B}_{ll}^{(quant,Bosons)} = \frac{\hbar}{2L_l} \sqrt{\frac{4\pi\hbar^2}{m} \left(\frac{n}{g_s \zeta\left(\frac{3}{2}\right)}\right)^{\frac{2}{3}}} \qquad (A.8)$$

Note that the argument of the square root is proportional to $k_B T_c$, the characteristic thermal energy at the transition point. Thus, for bosons, it is the energy scale associated with the critical condensation phenomenon, rather than a "Fermi"-type energy, that constitutes the dominant quantum contribution to the momentum variance.

**A.6. Synthesis and Unified Form**

The results from equations (A.7) and (A.8), obtained in the extreme quantum limits, share a common structure: $B_{ll}$ is proportional to $1/L_l$ and to the square root of a quantity proportional to $n^{2/3}$. These quantum terms, denoted $\Delta_F$ and $\Delta_B$, represent the fermionic degeneracy pressure and the bosonic condensation energy scale, respectively.

These specific expressions motivate the unified and interpolating form presented in equation (16) and (17) of the main text. This form, $\mathcal{B}_{ll} = \frac{\hbar}{2L_l}\sqrt{2\pi m k_B T + \Delta(n)}$, seamlessly captures the continuous transition between the classical regime, dominated by thermal motion (first term under the root), and the quantum degenerate regime, dominated by quantum statistics (second term $\Delta(n)$).